\documentclass[12pt]{article}

\usepackage{amsfonts,amssymb,bm,cite,amsmath}
\usepackage[dvips]{graphicx}
\usepackage[dvips]{psfrag}

\newcommand {\beq}{\begin{equation}}
\newcommand {\eeq}{\end{equation}}
\newcommand {\beqa}{\begin{eqnarray}}
\newcommand {\eeqa}{\end{eqnarray}}

\newcommand {\I}{{\scriptscriptstyle [I]}}
\newcommand {\Iprime}{{\scriptscriptstyle [I']}}
\newcommand {\IIprime}{{\scriptscriptstyle [II']}}

\allowdisplaybreaks

\begin{document}
\setlength{\oddsidemargin}{0cm}
\setlength{\baselineskip}{7mm}

\begin{titlepage}
\renewcommand{\thefootnote}{\fnsymbol{footnote}}
\begin{normalsize}
\begin{flushright}
\begin{tabular}{l}
OU-HET 575\\
March 2007
\end{tabular}
\end{flushright}
  \end{normalsize}

~~\\

\vspace*{0cm}
    \begin{Large}
       \begin{center}
        {T-duality, Fiber Bundles and Matrices}
       \end{center}   
    \end{Large}
\vspace{1cm}

\begin{center}
           Takaaki I{\sc shii}\footnote
            {
e-mail address : 
ishii@het.phys.sci.osaka-u.ac.jp},
           Goro I{\sc shiki}\footnote
            {
e-mail address : 
ishiki@het.phys.sci.osaka-u.ac.jp},
Shinji S{\sc himasaki}\footnote
            {
e-mail address : 
shinji@het.phys.sci.osaka-u.ac.jp}
           {\sc and}
           Asato T{\sc suchiya}\footnote
           {
e-mail address : tsuchiya@het.phys.sci.osaka-u.ac.jp}\\
      \vspace{1cm}
                    
               {\it Department of Physics, Graduate School of  
                     Science}\\
               {\it Osaka University, Toyonaka, Osaka 560-0043, Japan}
               
\end{center}

\vspace{1.5cm}

\begin{abstract}
\noindent
We extend the T-duality for gauge theory to that 
on curved space described as a nontrivial fiber bundle.
We also present a new viewpoint concerning
the consistent truncation and 
the T-duality for gauge theory
and discuss the relation between the vacua on the total space and on the
base space. 
As examples, we consider $S^3(/Z_k)$, $S^5(/Z_k)$ and
the Heisenberg nilmanifold.
\end{abstract}
\vfill
\end{titlepage}
\vfil\eject

\setcounter{footnote}{0}


\section{Introduction}
\setcounter{equation}{0}
\renewcommand{\thefootnote}{\arabic{footnote}} 
Emergence of space-time is 
one of the key concepts in nonperturbative definition
of superstring or M-theory by matrix models \cite{BFSS,IKKT}.
This phenomenon in field theory was found over two decades ago
in the large $N$ reduction of gauge theories 
\cite{EK,Parisi:1982gp,Bhanot:1982sh,Gross:1982at,Das:1982ux}, which
states equivalence under some conditions between a large $N$  
gauge theory and 
the matrix model that is its dimensional reduction to a point.
This equivalence originates from the fact 
that the eigenvalues of matrices can be interpreted as momenta.
This interpretation reappeared in  
the T-duality between the low-energy effective theories for D$p$-branes 
and for D$(p-1)$-branes \cite{Taylor:1996ik,Taylor:1997dy}.
More concretely, this T-duality tells that 
a $U(N)$ gauge theory on $R^p\times S^1$ is equivalent to
the $U(N\times\infty)$ gauge theory that is its dimensional reduction 
to $R^p$ if 
a periodicity condition is imposed to the theory on $R^p$. 

The main purpose of this paper is to extend the T-duality for gauge theory to
that on curved space described as a nontrivial fiber bundle.
The above mentioned T-duality
is concerning a trivial $S^1$ bundle, $R^p\times S^1$. 
We restrict ourselves to principal $S^1$
bundles and show the T-duality between the gauge theories on the total
space and on the base space. We also present a new viewpoint
concerning the consistent truncation and
the T-duality for gauge theory. 
Furthermore, we discuss the properties of
the vacua\footnote{Throughout
this paper, we consider gauge theories on Riemannian manifolds with
a positive-definite metric. In the following arguments, we can
easily add the time direction as direct product. To be precise, the 
`vacua' in this paper mean the classical vacua of the corresponding
gauge theories on this direct product space.}
on the total space and the base space.
In our previous publication \cite{ISTT} on the gauge/gravity correspondence
for the $SU(2|4)$ symmetric theories 
\cite{Lin:2005nh} (see also 
\cite{ITT,Ling:2006up,Lin:2006tr,Hikida:2006qb,Ling:2006xi,
vanAnders:2007ky}), we showed the T-duality between 
$\mathcal{N}=4$ super Yang Mills (SYM) on $R\times S^3(/Z_k)$ and $2+1$
SYM on $R\times S^2$, which is suggested from the gravity
side. This is regarded as the T-duality on 
$S^3(/Z_k)$, which is a nontrivial $S^1$ fibration over $S^2$. 
In this paper, we generalize this result.
Our findings would be useful for the study of describing curved 
space-time in matrix models \cite{Hanada:2005vr,Hanada:2006gg,Furuta:2006kk}
as well as the study of curved D-branes.
															
This paper is organized as follows. In section 2, we review
the T-duality for gauge theory in a standard way. In section 3, we present
a new viewpoint concerning the consistent truncation and the 
T-duality for gauge theory. Although this 
viewpoint is not necessarily needed for the proof of the T-duality 
on fiber bundle, it is interesting itself and 
indeed makes the T-duality for gauge theory more plausible.
In section 4, we 
consider a dimensional reduction from the total space of a principal $S^1$
bundle to its base space. In section 5, we show the T-duality between
the gauge theories on the base space and the total space. 
In section 6, we discuss the properties of  
the nontrivial vacua on the total space and the base space.
We classify the vacua on the total space and discuss their relation to
the vacua on the base space.
In section 7, we present 
some examples: $S^3(/Z_k)$, $S^5(/Z_k)$ and the Heisenberg nilmanifold.
Section 8 is devoted to conclusion and
discussion.


\section{Review of T-duality for gauge theory}
\setcounter{equation}{0}
In this section, we give a standard review of the T-duality between
the gauge theories on $R^p \times S^1$ and $R^p$
\cite{Taylor:1996ik,Taylor:1997dy}. 
We first consider pure Yang Mills on $R^p \times S^1$:
\begin{eqnarray}
S_{p+1}=\frac{1}{g_{p+1}^2}\int d^{p+1}z \:\frac{1}{4}
\mbox{Tr}(F_{MN}F_{MN}),
\label{S_p+1}
\end{eqnarray}
where $z^{M}\;(M=1,\cdots,p+1)$ are decomposed into 
$(x^{\mu},y)\; (\mu=1,\cdots,p)$, $y$ parameterizes $S^1$ with the radius 
$R$ and $F_{MN}=\partial_MA_N-\partial_NA_M-i[A_M,A_N]$.
By putting $A_{\mu}=a_{\mu},\;A_y=\phi$ and dropping the $y$-dependence,
this theory is 
dimensionally reduced to 
Yang Mills with a Higgs field $\phi$ on $R^p$:
\begin{eqnarray}
S_p=\frac{1}{g_p^2}\int d^px \: 
\mbox{Tr}\left( \frac{1}{4}f_{\mu\nu}f_{\mu\nu}
+\frac{1}{2}D_{\mu}\phi D_{\mu}\phi\right),
\label{S_p}
\end{eqnarray}
where 
$f_{\mu\nu}=\partial_{\mu}a_{\nu}-\partial_{\nu}a_{\mu}-i[a_{\mu},a_{\nu}]$,
$D_{\mu}\phi=\partial_{\mu}\phi-i[a_{\mu},\phi]$ and 
$g_p^2=\frac{1}{2\pi R}g_{p+1}^2$. Here, 
we adopt $U(N \times \infty)$ as the gauge group of $S_p$.
Namely, the fields in $S_p$
are hermitian matrices consisting of 
infinitely many blocks, each of which is an $N\times N$ matrix. We label
the blocks by $(s,t)$, where $s,\:t$ run from $-\infty$ to $\infty$.
Then, $S_p$ is expressed in terms of the blocks as follows:
\begin{eqnarray}
S_p
=\frac{1}{g_p^2}\int d^px \: 
\sum_{s,t}\mbox{tr}\left( \frac{1}{4}f_{\mu\nu}^{(s,t)}f_{\mu\nu}^{(t,s)}
+\frac{1}{2}(D_{\mu}\phi)^{(s,t)}(D_{\mu}\phi)^{(t,s)}\right),
\label{S_p in terms of blocks}
\end{eqnarray}
where tr stands for the trace over the $N\times N$ matrix.

We make an $S^1$ compactification with the radius $\tilde{R}$ 
in the $\phi$ direction by imposing
the following conditions on the fields:
\begin{eqnarray}
&&Ua_{\mu} U^{\dagger}=a_{\mu}, \nonumber\\
&&U\phi U^{\dagger}=\phi+2\pi \tilde{R} \mathbf{1}_{N\times\infty},
\end{eqnarray}
where $U$ is the `shift' matrix with infinite size,
\begin{eqnarray}
U=
\begin{pmatrix}
 \ddots &\ddots & & & &\\
  &\mathbf{0}_N &\mathbf{1}_N & & &\\
  & &\mathbf{0}_N &\mathbf{1}_N& & \\
  & & &\mathbf{0}_N&\mathbf{1}_N&  \\
  & & & &\mathbf{0}_N&\ddots       \\
  & & & & &\ddots& 
\end{pmatrix}.
\end{eqnarray}
These conditions are expressed in terms of the block components as 
\begin{eqnarray}
&&a_{\mu}^{(s+1,t+1)}=a_{\mu}^{(s,t)}, \nonumber\\
&&\phi^{(s+1,t+1)}=\phi^{(s,t)}+2\pi \tilde{R} \delta_{st}\mathbf{1}_N.
\end{eqnarray}
They can be solved as 
\begin{eqnarray}
a_{\mu}=\hat{a}_{\mu}+\tilde{a}_{\mu},\;\;\;\phi=\hat{\phi} + \tilde{\phi}
\end{eqnarray}
with
\begin{eqnarray}
\hat{a}_{\mu}=0, \;\;\;
\hat{\phi}=2\pi \tilde{R} \:
\mbox{diag}(\cdots,s-1,s,s+1,\cdots)\otimes \mathbf{1}_N 
\;\;\;(\hat{\phi}^{(s,t)}=2\pi \tilde{R}s\delta_{st}) 
\label{background}
\end{eqnarray}
and
\begin{eqnarray}
\tilde{a}_{\mu}^{(s,t)}=\tilde{a}_{\mu}^{(s-t)}, \;\;\;
\tilde{\phi}^{(s,t)}=\tilde{\phi}^{(s-t)}.
\label{periodicity}
\end{eqnarray}
The background (\ref{background}) is a vacuum of (\ref{S_p}).
The fluctuations around the vacuum, $\tilde{a}_{\mu}^{(s,t)}$ and 
$\tilde{\phi}^{(s,t)}$,
depend only on $s-t$ as indicated in (\ref{periodicity}), which
represents a periodicity.
The above procedure should be called orbifolding.

By making the Fourier transformation, which turns out to be interpreted as 
the T-duality, one can recover pure Yang Mills on $R^p\times S^1$, 
where the radius of the
original $S^1$
is $R$. The fields on $R^p \times S^1$ are defined in terms of 
the fields on $R^p$ as 
\begin{eqnarray}
&&A_{\mu}(x,y)=\sum_w \tilde{a}_{\mu}^{(w)}(x) 
e^{-\frac{i}{R}wy}, \nonumber\\
&&A_y(x,y)=\sum_w \tilde{\phi}^{(w)}(x) e^{-\frac{i}{R}wy}.
\label{Fourier transformation}
\end{eqnarray}
The radius of the original $S^1$, $R$,
is related to the radius of the dual $S^1$,
$\tilde{R}$, as 
\begin{eqnarray}
R=\frac{1}{2\pi\tilde{R}}.
\label{relation between radii}
\end{eqnarray}
Then, the block components in (\ref{S_p in terms of blocks}) are evaluated as
\begin{align}
(D_{\mu}\phi(x))^{(s,t)}
&=\partial_{\mu}\tilde{\phi}^{(s-t)}(x)
+i2\pi \tilde{R}(s-t)\tilde{a}_{\mu}^{(s-t)}(x) \nonumber\\
&\;\;\;\;\;-i\sum_u(\tilde{a}_{\mu}^{(s-u)}(x)\tilde{\phi}^{(u-t)}(x)
-\tilde{\phi}^{(s-u)}(x)\tilde{a}_{\mu}^{(u-t)}(x)) 
\nonumber\\
&=\frac{1}{2\pi R}\int_0^{2\pi R}dy 
F_{\mu y}(x,y)e^{\frac{i}{R}(s-t)y}, \nonumber\\
f_{\mu\nu}^{(s,t)}(x)&=
\frac{1}{2\pi R}\int_0^{2\pi R}dy 
F_{\mu\nu}(x,y)e^{\frac{i}{R}(s-t)y}.
\label{(s,t)}
\end{align}
Substituting (\ref{(s,t)}) into
(\ref{S_p in terms of blocks}) yields
\begin{eqnarray}
S_p=\frac{1}{g_p^2}\frac{1}{2\pi R}\sum_w\int d^{p+1}z \: 
\frac{1}{4}\mbox{tr}(F_{MN}F_{MN}).
\label{S_p2}
\end{eqnarray}
By dividing the above expression by the overall factor $\sum_w$, which gives
an infinite constant, 
one indeed reproduces the original pure Yang Mills on
$R^p \times S^1$ (\ref{S_p+1}) with the gauge group $U(N)$. 

In the context of the D-brane effective theories, the above procedure is
interpreted as the T-duality
between D$p$-brane and D$(p-1)$-brane, although the $9-p$ Higgs fields and
the fermions are omitted here for simplicity.
The background (\ref{background})
represents an infinite array of stacks of $N$ 
coincident D$(p-1)$-branes, where `$s$' labels the $s$-th stack. 
The distance
between the neighboring stacks is $2\pi \tilde{R}$. (\ref{periodicity}) 
expresses the periodicity which produces the dual $S^1$ with the radius 
$\tilde{R}$.
$\tilde{a}_{\mu}^{(w)}$ and $\tilde{\phi}^{(w)}$ 
represent an open string stretched between
the $s$-th stack and the $(s+w)$-th stack, so that $-w$ corresponds to the 
winding number around the dual $S^1$. 
In (\ref{Fourier transformation}), the winding number $-w$ is reinterpreted as
the momentum $-w/R$ along the original $S^1$ 
with the radius $R$. The relation between the radii 
(\ref{relation between radii}) is the same as that for the T-duality in 
string theory.
Dividing (\ref{S_p2}) by the overall factor
$\sum_w$ corresponds to 
extracting a single period. In this way, the effective theory for 
a stack of $N$ coincident D$p$-branes is obtained through the T-duality.

\section{Consistent truncation and T-duality}
\setcounter{equation}{0}
In the previous section, we reviewed the T-duality for gauge theory 
in a standard way.
In this section, we present a new viewpoint 
concerning the consistent truncation and the T-duality.

Let the gauge group in (\ref{S_p+1}) be $U(M)$.
We consider a pure-gauge background,
\begin{eqnarray}
&&\hat{A}_{\mu}=0=-i\partial_{\mu}VV^{\dagger},\nonumber\\ 
&&\hat{A}_{y}
=\frac{1}{R}\mbox{diag}(\cdots,\underbrace{n_{s-1},\cdots,n_{s-1}}_{N_{s-1}},
\underbrace{n_s,\cdots,n_s}_{N_s},\underbrace{n_{s+1},\cdots,n_{s+1}}_{N_{s+1}}
,\cdots)
=-i\partial_{y}VV^{\dagger},
\label{background of YM on R^pxS^1}
\end{eqnarray}
with
\begin{eqnarray}
V=\mbox{diag}(\cdots,
\underbrace{e^{\frac{i}{R}n_{s-1}y},\cdots,e^{\frac{i}{R}n_{s-1}y}}_{N_{s-1}},
\underbrace{e^{\frac{i}{R}n_sy},\cdots,e^{\frac{i}{R}n_sy}}_{N_s},
\underbrace{e^{\frac{i}{R}n_{s+1}y},\cdots,e^{\frac{i}{R}n_{s+1}y}}_{N_{s+1}},
\cdots),
\label{V}
\end{eqnarray}
where $M=\cdots + N_{s-1}+N_s+N_{s+1}+\cdots$.
Due to the single-valuedness of $V$, all $n_s$ must be integers.
We assume that all $n_s$ are different.
This background naturally induces a block structure for $M\times M$ matrices.
We label the blocks by $(s,t)$,
where the $(s,t)$ block is an $N_s\times N_t$ matrix.

We denote the fluctuations of $A_M$ around 
the background (\ref{background of YM on R^pxS^1}) by $\tilde{A}_M$,
while we continue to use $A_M$ for the fields around the trivial 
background $A_M=0$.
Since the background (\ref{background of YM on R^pxS^1}) is gauge-equivalent to
the trivial background, we have a relation
\begin{eqnarray}
A_M=-i\partial_MV^{\dagger}V+V^{\dagger}(\hat{A}_M+\tilde{A}_M)V,
\end{eqnarray}
which is equivalent to 
\begin{eqnarray}
A_M=V^{\dagger}\tilde{A}_MV.
\label{relation between A_M^(0) and tildeA_M}
\end{eqnarray}
For the $(s,t)$ block, (\ref{relation between A_M^(0) and tildeA_M})
implies 
\begin{eqnarray}
A_M^{(s,t)}=e^{-\frac{i}{R}(n_s-n_t)y}\tilde{A}_M^{(s,t)}.
\label{relation between blocks}
\end{eqnarray}

We make the Fourier expansions for $A_M^{(s,t)}$ and $\tilde{A}_M^{(s,t)}$
with respect to the $y$ direction as 
\begin{eqnarray}
&&A_M^{(s,t)}(x,y)
=\sum_mA_{M,m}^{(s,t)}(x)e^{\frac{i}{R}my}, \nonumber\\
&&\tilde{A}_M^{(s,t)}(x,y)
=\sum_m\tilde{A}_{M,m}^{(s,t)}(x)e^{\frac{i}{R}my}.
\end{eqnarray}
$\mbox{}$From (\ref{relation between blocks}), we obtain a relation between
the Kaluza-Klein (KK) modes,
\begin{eqnarray}
A_{M,m-(n_s-n_t)}^{(s,t)}(x)=\tilde{A}_{M,m}^{(s,t)}(x).
\label{relation between modes}
\end{eqnarray}
The theory around the trivial background of (\ref{S_p+1}) is written
in terms of $A_{M,m}^{(s,t)}$ while the theory around
the background (\ref{background of YM on R^pxS^1}) of (\ref{S_p+1})
in terms
of $\tilde{A}_{M,m}^{(s,t)}$. The two theories are equivalent under the 
identification (\ref{relation between modes}).
The trivial background $A_M=0$ corresponds to the trivial vacuum of the theory.
Due to the variety of the choices of $n_s$ and $N_s$, 
we have many different representations 
of the theory around the trivial vacuum.



In the usual KK reduction, one keeps only $A_{M,0}^{(s,t)}$
in the theory around the trivial background of (\ref{S_p+1}). This is 
a consistent truncation, because the momentum `$m$' is conserved, and 
one obtains the theory around the
trivial vacuum $a_{\mu}=0,\;\phi=0$ of (\ref{S_p}). 
Similarly,  
one can keep only $\tilde{A}_{M,0}^{(s,t)}$ in the theory around 
the background (\ref{background of YM on R^pxS^1}) of (\ref{S_p+1}) 
to truncate (\ref{S_p+1}) consistently. 
It is seen from (\ref{background of YM on R^pxS^1}) that 
the resulting theory is the theory around a vacuum of (\ref{S_p}) given by
\begin{eqnarray}
\hat{a}_{\mu}=0,\;\;\;\hat{\phi}
=2\pi\tilde{R}\mbox{diag}(\cdots,\underbrace{n_{s-1},\cdots,n_{s-1}}_{N_{s-1}},
\underbrace{n_s,\cdots,n_s}_{N_s},\underbrace{n_{s+1},\cdots,n_{s+1}}_{N_{s+1}}
,\cdots).
\label{background of YM with Higgs on R^p}
\end{eqnarray}
In this theory, $\tilde{A}_{\mu,0}^{(s,t)}$ and $\tilde{A}_{y,0}^{(s,t)}$
are identified with $\tilde{a}_{\mu}^{(s,t)}$ and $\tilde{\phi}^{(s,t)}$, 
respectively.
This theory is no longer equivalent to the theory around the trivial vacuum of 
(\ref{S_p}), although these two theories originate from the same theory.
In other words, we can obtain many different 
theories by consistently truncating the original
theory in different ways. 
Indeed, (\ref{relation between modes}) tells us that
keeping only $A_{M,-(n_s-n_t)}^{(s,t)}$ in the theory 
around the trivial background of (\ref{S_p+1})
yields the theory around 
the vacuum (\ref{background of YM with Higgs on R^p}) of (\ref{S_p}).
That this is a consistent truncation can also be understood from the fact that
the sum of the charge `$n_s-n_t$' and the momentum `$m$' is conserved 
because so is each of them.
Note that in the theory around the vacuum 
(\ref{background of YM with Higgs on R^p}) of (\ref{S_p})
the gauge symmetry $U(M)$ is spontaneously broken to
$\cdots \times U(N_{s-1})\times U(N_s) \times U(N_{s+1}) \times \cdots$.

By using the above discussions, 
we can easily show in an alternative way
the T-duality reviewed in the previous section.
Let us consider the case in which $M=N\times \infty$, 
$s$ runs from $-\infty$ to
$\infty$, $N_s=N$ for all $s$ and $n_s=s$. 
In this case, the vacuum (\ref{background of YM with
Higgs on R^p}) is nothing but the vacuum (\ref{background})
considered in the previous
section. In the theory around the trivial background of (\ref{S_p+1}),
we impose the constraint
\begin{eqnarray}
A_{M,m}^{(s,t)}=A_{M,m}^{(s-t)},
\end{eqnarray}
and keep only $A_{M,-(s-t)}^{(s-t)}$. 
The summations over the block indices $s,t,\cdots$ are identified with
the summations over the momenta. From the momentum conservation, we
have the overall factor $\sum_w$.
Thus we obtain
the theory around the trivial vacuum
of $U(N)$ Yang Mills on $R^p\times S^1$
with the overall factor $\sum_w$,
where $A_{M,m}^{(-m)}$ is identified with the KK mode 
$A_{M,m}$ of the $U(N)$ theory.
We see, therefore, from the discussion in the previous paragraph that
the theory around the vacuum (\ref{background}) of (\ref{S_p}) with the 
periodicity condition (\ref{periodicity}) imposed is equivalent to the theory
around the trivial vacuum of 
(\ref{S_p+1}) with the gauge group $U(N)$ and the overall factor $\sum_w$.
This is indeed the T-duality reviewed in the previous section.

\section{Dimensional reduction from total space to base space}
\setcounter{equation}{0}
In this section, we perform a dimensional reduction from the total space
of a principal $S^1$ bundle to its base space.
We consider a principal $S^1$ bundle whose total space 
is a $(D+1)$-dimensional manifold $P$ and whose base space is
a $D$-dimensional manifold $B$. The projection is given by 
$\pi : P\rightarrow B$. The base space $B$ has a covering 
$\{U_{\I}\}\;\;(I=1,2,\cdots)$, each element of 
which is parameterized by $x^{\mu}_{\I}\;\;(\mu=1,\cdots,D)$.
The total space $P$ has a covering $\{\pi^{-1}(U_{\I})\}$.
$\pi^{-1}(U_{\I})$ is diffeomorphic to $U_{\I}\times S^1$ by the local 
trivialization, so that it is parameterized by 
$z^M_{\I}=(x^{\mu}_{\I},y_{\I})\;\;(M=1,\cdots,D+1)$,
where $y_{\I}=z^{D+1}_{\I}$ parameterizes the $S^1$ and $0\leq y_{\I} <2\pi R$.
If there is overlap between $U_{\I}$ and $U_{\Iprime}$, the relation
between $y_{\I}$ and $y_{\Iprime}$ is determined by the transition
function $e^{-\frac{i}{R}v_{\IIprime}}$ 
as $y_{\Iprime}=y_{\I}-v_{\IIprime}(x_{\I})$.
In the following, we add a subscript or superscript $[I]$ to quantities
which are evaluated on $U_{\I}$.
Quantities without such a subscript or superscript are 
independent of which patch is used to evaluate them.

We assume that the total space possesses the $U(1)$ isometry
in the fiber direction and the size of the fiber, namely the radius of $S^1$,
is constant. The metrics that satisfy such conditions generally take the form
on $\pi^{-1}(U_{\I})$
\begin{eqnarray}
ds_{D+1}^2=G_{MN}^{\I}dz^M_{\I}dz^N_{\I}
=g_{\mu\nu}^{\I}(x_{\I})dx^{\mu}_{\I}dx^{\nu}_{\I}
+(dy_{\I}+b_{\mu}^{\I}(x_{\I})dx^{\mu}_{\I})^2,
\label{metric}
\end{eqnarray}
where $b^{\I}=b_{\mu}^{\I}dx^{\mu}_{\I}$ 
must be transformed as $b^{\Iprime}=b^{\I}+dv_{\IIprime}$.
$\mbox{}$From this metric, one can define a connection 1-form 
on the principal bundle as follows.
First, note that
connection 1-forms in general take the form
\begin{eqnarray}
\omega=\frac{1}{R}(dy_{\I}+t_{\mu}^{\I}(x_{\I})dx^{\mu}_{\I}),
\end{eqnarray}
where $t^{\I}$ must be transformed as $t^{\Iprime}=t^{\I}+dv_{\IIprime}$. 
Second, we introduce an orthonormal basis for
the tangent space of the total space, $E_{A}\;\;(A=1,\cdots,D+1)$, such
that the direction of $E_{D+1}$ coincides with the fiber direction.
Explicitly, the elements of $E_A$ are given by
\begin{eqnarray}
&&E_{\alpha}^{\I\mu}=e_{\alpha}^{\I\mu},\;\;\;
E_{\alpha}^{\I y}=-e_{\alpha}^{\I\nu}b_{\nu}^{\I}=-b_{\alpha}^{\I}, \nonumber\\
&&E_{D+1}^{\I\mu}=0, \;\;\;E_{D+1}^{\I y}=1,
\label{orthonormal basis}
\end{eqnarray}
where $\alpha=1,\cdots,D$ and $e_{\alpha}^{\I\mu}$ is determined from
$g^{\I\mu\nu}=e_{\alpha}^{\I\mu}e_{\alpha}^{\I\nu}$.
$E_{\alpha}$ span the subspace orthogonal to the fiber direction.
Then, $\omega$ is determined from the condition $\omega(E_{\alpha})=0$ 
for all $\alpha$ as 
\begin{eqnarray}
\omega=\frac{1}{R}(dy_{\I}+b^{\I}).
\end{eqnarray}
The orthonormal basis $E^A$ of the cotangent space
dual to (\ref{orthonormal basis}) is given by
\begin{eqnarray}
&&E_{\mu}^{\I\alpha}=e_{\mu}^{\I\alpha}, \;\;\; E_y^{\I\alpha}=0, \nonumber\\
&&E_{\mu}^{\I D+1}=b_{\mu}^{\I},\;\;\;E_y^{\I D+1}=1,
\label{vielbeins}
\end{eqnarray}
which are identified with the vielbeins of the total space. The indices `$A$' 
are the local
Lorentz indices for the total space.
One can identify the space spanned by $E_{\alpha}$, in which the inner product
is given by $G$ in (\ref{metric}),
with the tangent space  
of the base space with the same inner product. 
Then, it follows from (\ref{vielbeins}) that 
$e_{\mu}^{\I\alpha}$ are the vielbeins of the $D$-dimensional
base space, namely $g_{\mu\nu}^{\I}$ are the metric of 
the base space and $\alpha$
are the local Lorentz indices for the base space.
Note that $\frac{1}{R}b_{\mu}^{\I}$ gives a connection 
1-form of the vector bundle associated with the principal bundle.
The spin connections, 
$\Omega_{A\;\;C}^{\;\;\:B}=E_A^{\I M}\Omega_{M\;\;C}^{\I B}$, 
are determined from (\ref{vielbeins}) as
\begin{eqnarray}
&&\Omega_{\alpha \;\; \gamma}^{\;\;\beta}
=\omega_{\alpha \;\; \gamma}^{\;\;\beta},\;\;\;
\Omega_{\alpha\;\;D+1}^{\;\;\beta}=\frac{1}{2}b_{\alpha\beta}, \nonumber\\
&&\Omega_{D+1\;\;\beta}^{\;\;\;\;\;\;\;\alpha}
=\frac{1}{2}b_{\beta\alpha},\;\;\;
\Omega_{D+1\;\;D+1}^{\;\;\;\;\;\;\;\alpha}=0,
\label{spin connection}
\end{eqnarray}
where $\omega_{\alpha \;\; \gamma}^{\;\;\beta}$ are the spin connections on
the base space evaluated from $e_{\mu}^{\I\alpha}$, and
$b_{\alpha\beta}=\nabla_{\alpha}b_{\beta}^{\I}-\nabla_{\beta}b_{\alpha}^{\I}$.

We start with pure Yang Mills on the $(D+1)$-dimensional total space:
\begin{align}
S_{D+1}
=\frac{1}{g_{D+1}^2}\int d^{D+1}z \:\sqrt{G}\frac{1}{4}
\mbox{Tr}(F_{AB}F_{AB}),
\label{S_D+1}
\end{align}
where
$d^{D+1}z \:\sqrt{G}$ represents the invariant volume.
We dimensionally reduce this theory to Yang Mills with a Higgs field on
the $D$-dimensional base space. 
Since we decomposed the (co)tangent space of the total space into the fiber 
direction and the directions orthogonal to it in
(\ref{orthonormal basis}) and (\ref{vielbeins}), we naturally
relate the gauge fields $A_A$ on
the total space to the gauge fields $a_{\alpha}$ and the Higgs field $\phi$ 
on the base space as follows:
\begin{eqnarray}
&&A_{\alpha}=a_{\alpha}, \nonumber\\
&&A_{D+1}=\phi,
\label{reduction}
\end{eqnarray}
where we assume that the both sides in (\ref{reduction}) 
are independent of $y_{\I}$.
By using (\ref{spin connection}), we evaluate the field strength on the 
total space as 
\begin{eqnarray}
&&F_{\alpha\beta}=f_{\alpha\beta}+b_{\alpha\beta}, \nonumber\\
&&F_{\alpha D+1}=D_{\alpha}\phi,
\label{field strength on total space 2}
\end{eqnarray}
where $f_{\alpha\beta}=\nabla_{\alpha}a_{\beta}-\nabla_{\beta}a_{\alpha}
-i[a_{\alpha},a_{\beta}]$.
By using (\ref{field strength on total space 2}) and 
$\sqrt{G^{\I}}=\sqrt{g^{\I}}$,          
we obtain from (\ref{S_D+1})
Yang Mills with the Higgs field $\phi$ 
on the base space\footnote{This 
action is formally the same as that derived in \cite{Cremmer}, where
the compactification of gravitational and Yang Mills system from a direct
product
space-time $M\times S^1$ to $M$ is considered, and $b_{\alpha}$
represents fluctuation of the metric on $M\times S^1$.}:
\begin{align}
S_D
= \frac{1}{g_D^2}\int d^Dx \sqrt{g} \:\mbox{Tr}
\left(\frac{1}{4}
(f_{\alpha\beta}+b_{\alpha\beta}\phi)(f_{\alpha\beta}+b_{\alpha\beta}\phi)
+\frac{1}{2}D_{\alpha}\phi D_{\alpha}\phi \right),
\label{S_D}
\end{align}
where $g_D^2=\frac{1}{2\pi R}g_{D+1}^2$.
Note that there appears the $U(1)$ curvature $b_{\alpha\beta}$ in (\ref{S_D})

\section{T-duality on fiber bundle}
\setcounter{equation}{0}
The discussion on the consistent truncation of Yang Mills on the total space
of the principal $S^1$ bundle
proceeds parallel to that of Yang Mills on $R^p\times S^1$ in section 3.
By using the discussion, we can show the T-duality between the gauge theories
on the total space
and on the base space.
As before, let the gauge group in (\ref{S_D+1}) be $U(M)$.
We consider a gauge transformation which 
is an analogue of $V$ in (\ref{V}).
Such a gauge transformation should be defined locally on each patch.
It is given on $\pi^{-1}(U_{\I})$ by
\begin{align}
V_{\I}=\mbox{diag}(\cdots,
\underbrace{e^{\frac{i}{R}n_{s-1}y_{\I}},\cdots,
e^{\frac{i}{R}n_{s-1}y_{\I}}}_{N_{s-1}},
\underbrace{e^{\frac{i}{R}n_{s}y_{\I}},\cdots,
e^{\frac{i}{R}n_{s}y_{\I}}}_{N_s},
\underbrace{e^{\frac{i}{R}n_{s+1}y_{\I}},
\cdots,e^{\frac{i}{R}n_{s+1}y_{\I}}}_{N_{s+1}},\cdots),
\label{VI}
\end{align}
where $M=\cdots +N_{s-1}+N_s+N_{s+1}+\cdots$.
Here all $n_s$ are different and 
must be integers due to the single-valuedness of $V_{\I}$.
$\mbox{}$From (\ref{VI}), we can evaluate the pure-gauge background 
on $\pi^{-1}(U_{\I})$ as 
\begin{align}
&\hat{A}_{\alpha}^{\I}
=-iE^{\I M}_{\alpha}\frac{\partial V_{\I}}{\partial z^M_{\I}}V_{\I}^{\dagger}
=-\frac{1}{R}b_{\alpha}^{\I}\:
\mbox{diag}(\cdots,
\underbrace{n_{s-1},\cdots,n_{s-1}}_{N_{s-1}},
\underbrace{n_s,\cdots,n_s}_{N_s},
\underbrace{n_{s+1},\cdots,n_{s+1}}_{N_{s+1}},\cdots),\nonumber\\
&\hat{A}_{D+1}
=-iE^{\I M}_{D+1}\frac{\partial V_{\I}}{\partial z^M_{\I}}V_{\I}^{\dagger}
=\frac{1}{R}\mbox{diag}
(\cdots,
\underbrace{n_{s-1},\cdots,n_{s-1}}_{N_{s-1}},
\underbrace{n_s,\cdots,n_s}_{N_s},
\underbrace{n_{s+1},\cdots,n_{s+1}}_{N_{s+1}},\cdots).
\label{pure gauge background}
\end{align}
Note that $\hat{A}_{\alpha}^{\I}$ is patch-dependent 
as $b_{\alpha}^{\I}$ does. 
This patch-dependence originates from
considering a particular background.
If there is overlap between $U_{\I}$ and $U_{\Iprime}$, 
$\hat{A}_{\alpha}^{\I}$ 
is gauge-transformed to $\hat{A}_{\alpha}^{\Iprime}$ by 
\begin{align}
V_{\IIprime}
&=V_{\Iprime}V_{\I}^{\dagger} \nonumber\\
&=\mbox{diag}(\cdots,
\underbrace{e^{-\frac{i}{R}n_{s-1}v_{\IIprime}},\cdots,
e^{-\frac{i}{R}n_{s-1}v_{\IIprime}}}_{N_{s-1}},
\underbrace{e^{-\frac{i}{R}n_{s}v_{\IIprime}},\cdots,
e^{-\frac{i}{R}n_{s}v_{\IIprime}}}_{N_s}, \nonumber\\
&\;\;\;\;\;\;\;\;\;\;\underbrace{e^{-\frac{i}{R}n_{s+1}v_{\IIprime}},\cdots,
e^{-\frac{i}{R}n_{s+1}v_{\IIprime}}}_{N_{s+1}},\cdots).
\label{gauge transformation}
\end{align}
while $\hat{A}_{D+1}$ is invariant.
$e^{-\frac{i}{R}v_{\IIprime}}$ is nothing
but the transition function between $U_{\I}$ and $U_{\Iprime}$, so that
$V_{\IIprime}$ is well-defined.
The background (\ref{pure gauge background}) is gauge-equivalent to the 
trivial background $A_A=0$, which corresponds to
the trivial vacuum of the theory.

As in section 3, we denote the fluctuations on $\pi^{-1}(U_{\I})$
around the background
(\ref{pure gauge background}) by $\tilde{A}_A^{\I}$,
while we continue to use $A_A$ for the gauge fields 
around the trivial background $A_{A}=0$, which are patch-independent.
The background (\ref{pure gauge background}) is gauge-transformed to
the trivial background by $V_{\I}^{\dagger}$, so that as in
(\ref{relation between blocks}) we have
\begin{eqnarray}
A_A^{(s,t)}=e^{-\frac{i}{R}(n_s-n_t)y_{\I}}
               \tilde{A}_A^{\I(s,t)}.
\label{relation for (s,t) block}
\end{eqnarray}
We also see from (\ref{gauge transformation})
\begin{eqnarray}
\tilde{A}_A^{\Iprime(s,t)}=e^{-\frac{i}{R}(n_s-n_t)v_{\IIprime}}
               \tilde{A}_A^{\I(s,t)}.
\label{relation for (s,t) block 2}
\end{eqnarray}
We can make the Fourier
transformations locally on each patch with respect to $y_{\I}$.
On $\pi^{-1}(U_{\I})$, $A_A^{(s,t)}$ and 
$\tilde{A}_A^{\I(s,t)}$ are expanded as
\begin{eqnarray}
&&A_A^{(s,t)}(x_{\I},y_{\I})=\sum_m A_{A,m}^{\I(s,t)}(x_{\I})
e^{\frac{i}{R}my_{\I}},  \nonumber\\
&&\tilde{A}_A^{\I(s,t)}(x_{\I},y_{\I})
=\sum_m \tilde{A}_{A,m}^{\I(s,t)}(x_{\I})e^{\frac{i}{R}my_{\I}}.
\end{eqnarray}
$\mbox{}$From these equalities, we easily see that 
\begin{eqnarray}
&&A_{A,m}^{\Iprime(s,t)}(x_{\Iprime})
=e^{\frac{i}{R}mv_{\IIprime}}A_{A,m}^{\I(s,t)}(x_{\I}), \nonumber\\
&&\tilde{A}_{A,m}^{\Iprime(s,t)}(x_{\Iprime})
=e^{\frac{i}{R}(m-(n_s-n_t))v_{\IIprime}}\tilde{A}_{A,m}^{\I(s,t)}(x_{\I}).
\label{transformation for KK modes}
\end{eqnarray}
The relation (\ref{relation for (s,t) block}) is translated to the relation
between the KK modes:
\begin{eqnarray}
A_{A,m-(n_s-n_t)}^{\I(s,t)}=
                             \tilde{A}_{A,m}^{\I (s,t)}.
\label{relation between KK modes}
\end{eqnarray}
This is of course consistent with (\ref{transformation for KK modes}).
The theory around the trivial background of (\ref{S_D+1}) is equivalent to
the theory around the background (\ref{pure gauge background}) 
of (\ref{S_D+1}) under the identification of the KK modes
(\ref{relation between KK modes}).

As in the $R^p\times S^1$ case, different consistent truncations
of the theory around the trivial vacuum of (\ref{S_D+1}) 
give rise to different theories on the base space.
The $U(1)$ isometry indeed
ensures that the following truncations are consistent ones.
Keeping only $A_{A,0}^{\I(s,t)}$ in the theory around the trivial background
of (\ref{S_D+1}) generates the theory around the trivial
vacuum $a_{\alpha}=0, \;\phi=0$ of (\ref{S_D}). 
Keeping only $\tilde{A}_{A,0}^{\I(s,t)}$ in the theory around the
background (\ref{pure gauge background}) is equivalent to keeping only
$A_{A,-(n_s-n_t)}^{\I(s,t)}$ in the theory around the trivial background,
and generates the theory around a nontrivial background of (\ref{S_D}), which
we will discuss shortly. By taking $M=N\times\infty$, $N_s=N$ and $n_s=s$ and 
imposing the periodicity $A_A^{(s,t)}=A_A^{(s-t)}$, 
the T-duality between the theories
on the total space and on the base space is achieved in the same way
as the $R^p\times S^1$ case.

It is seen from (\ref{reduction}) and (\ref{pure gauge background})
that keeping only $A_{A,-(n_s-n_t)}^{\I(s,t)}$ results in 
the theory around a background of (\ref{S_D}),
\begin{eqnarray}
&&\hat{a}_{\alpha}^{\I}
=-b_{\alpha}^{\I}\hat{\phi} ,\nonumber\\
&&\hat{\phi}=2\pi \tilde{R}\: \mbox{diag}
(\cdots,
\underbrace{n_{s-1},\cdots,n_{s-1}}_{N_{s-1}},
\underbrace{n_s,\cdots,n_s}_{N_s},
\underbrace{n_{s+1},\cdots,n_{s+1}}_{N_{s+1}},\cdots),
\label{nontrivial vacuum}
\end{eqnarray}
where $\tilde{R}=\frac{1}{2\pi R}$.
It is remarkable that the gauge fields take 
the monopole-like configuration described
by $b_{\alpha}^{\I}$.
We discuss the quantization of the fluxes in section 6.
The background (\ref{nontrivial vacuum}) would correspond to a vacuum of
(\ref{S_D}), because the background (\ref{pure gauge background}) 
corresponds to a vacuum of (\ref{S_D+1}). Indeed it satisfies
the equations
\begin{eqnarray}
&&\hat{f}_{\alpha\beta}+b_{\alpha\beta}\hat{\phi}=0, \nonumber\\
&&e_{\alpha}^{\mu}\partial_{\mu}\hat{\phi}
-i[\hat{a}_{\alpha},\hat{\phi}]=0,
\label{vacuum condition}
\end{eqnarray}
which give the conditions for the vacua.
If there is overlap between $U_{\I}$ and $U_{\Iprime}$,
the gauge fields and the Higgs field in $U_{\I}$ and $U_{\Iprime}$
are related by
the gauge transformation
\begin{eqnarray}
&&\hat{a}_{\alpha}^{\Iprime}
=-ie^{\I \mu}_{\alpha}\partial_{\mu}^{\I}V_{{\scriptscriptstyle [II']}}
   V_{\IIprime}^{\dagger}
  +V_{\IIprime}\hat{a}_{\alpha}^{\I}
   V_{\IIprime}^{\dagger}, \nonumber\\
&&\hat{\phi}=V_{\IIprime}\hat{\phi}
           V_{\IIprime}^{\dagger}.
\end{eqnarray}
We denote the $(s,t)$ block of fluctuations around (\ref{nontrivial vacuum}) 
on $U_{\I}$ by $\tilde{a}^{\I(s,t)}_{\alpha}$ and $\tilde{\phi}^{\I(s,t)}$,
which are identified with 
$\tilde{A}_{\alpha,0}^{\I(s,t)}=A_{\alpha,-(n_s-n_t)}^{\I(s,t)}$
and $\tilde{A}_{D+1,0}^{\I(s,t)}=A_{D+1,-(n_s-n_t)}^{\I(s,t)}$, 
respectively.
The fluctuations are gauge-transformed from $U_{\I}$ to $U_{\Iprime}$ as
\begin{align}
&\tilde{a}^{\Iprime(s,t)}_{\alpha}
=e^{-\frac{i}{R}(n_s-n_t)v_{\IIprime}}\tilde{a}_{\alpha}^{\I(s,t)},\nonumber\\
&\tilde{\phi}^{\Iprime(s,t)}
=e^{-\frac{i}{R}(n_s-n_t)v_{\IIprime}}\tilde{\phi}^{\I(s,t)}.
\label{gauge transf of fluctuations}
\end{align}

For completeness, we state explicitly the T-duality in this case:
the theory around (\ref{nontrivial vacuum}) of 
(\ref{S_D}) with $M=N\times\infty$, 
$N_s=N$, $n_s=s$ and the periodicity condition
$\tilde{a}_{\alpha}^{\I(s,t)}=\tilde{a}_{\alpha}^{\I(s-t)}$ and 
$\tilde{\phi}^{\I(s,t)}=\tilde{\phi}^{\I(s-t)}$ is equivalent to the theory
around the trivial vacuum of (\ref{S_D+1}) with the gauge group $U(N)$ and
the overall factor $\sum_w$.
The relation between the fields on the total space and on the base space is 
given by
\begin{eqnarray}
&&A_{\alpha}(x_{\I},y_{\I})=\sum_w \tilde{a}_{\alpha}^{{\I} (w)}(x_{\I})
e^{-\frac{i}{R}wy_{\I}}, \nonumber\\
&&A_{D+1}(x_{\I},y_{\I})=\sum_w \tilde{\phi}^{{\I} (w)}(x_{\I}) 
e^{-\frac{i}{R}wy_{\I}}.
\label{Fourier transf}
\end{eqnarray}
In order that
the fields in the lefthand sides in (\ref{Fourier transf}) are the ones
around the trivial vacuum of (\ref{S_D+1}), they must be
patch-independent. 
It is seen from (\ref{gauge transf of fluctuations}) 
they are indeed patch-independent. 
Interestingly, the monopole-like charges are 
identified with the momenta.
It is indeed easy to check explicitly that the Fourier transformation 
(\ref{Fourier transf})
realizes the T-duality, as we did
in section 2.

\section{Nontrivial vacua on total space}
\setcounter{equation}{0}
So far we have been concerned with the theory around the trivial vacuum on
the total space. In general, there are nontrivial vacua on the total space.
In this section, we discuss the nontrivial vacua on the total space and
their relation to the vacua on the base space.

First, we classify the vacua on the total space.
Let the gauge group of (\ref{S_D+1}) be $U(M)$.
The vacua of (\ref{S_D+1}) are given by the space  
of the flat connections modulo the gauge transformations, which are
parameterized by the holonomies (the Wilson lines) 
along the nontrivial generators of 
the fundamental group.  Let us consider the closed loop along
the fiber $S^1$. The Wilson line along the loop for the flat connections
is diagonalized as \cite{Hosotani:1983xw,Hosotani:1988bm}
\begin{align}
W&=P\exp\left(i\int_0^{2\pi R}dy_{\I}A_y(x_{\I},y_{\I})\right) \nonumber\\
&=\mbox{diag}
(\underbrace{e^{2\pi i\theta^{(1)}},\cdots,e^{2\pi i\theta^{(1)}}}_{M^{(1)}},
\cdots,
\underbrace{e^{2\pi i\theta^{(T)}},\cdots,e^{2\pi i\theta^{(T)}}}_{M^{(T)}}),
\label{Wilson line along S^1}
\end{align}
where $M=M^{(1)}+\cdots+M^{(T)}$, 
and all $\theta^{(a)}$ are  
constants different each other and  satisfying $0\leq \theta^{(a)} <1$.
If the loop is
contractable, $W=\mathbf{1}_M$, namely $T=1$ and $\theta^{(1)}=0$.
In the case of the nontrivial fiber bundles, $\theta^{(a)}$ are in general
discretized, as we will see shortly.
One can take a gauge in which $A_y$ is diagonal and constant:
\begin{align}
\hat{A}_y=\frac{1}{R}\mbox{diag}
(\underbrace{\theta^{(1)},\cdots,\theta^{(1)}}_{M^{(1)}},
\cdots,
\underbrace{\theta^{(T)},\cdots,\theta^{(T)}}_{M^{(T)}}),
\label{A_y}
\end{align}
which gives (\ref{Wilson line along S^1}).
By solving the flatness condition $F_{\mu y}^{\I}=0$, one finds
that $A_{\mu}^{\I}$ must have the same block structure as $\hat{A}_y$ and be 
$y_{\I}$-independent:
\begin{eqnarray}
\hat{A}_{\mu}^{\I}(x_{\I})
=\left(\begin{array}{ccc}
\hat{A}_{\mu}^{\I (1)}(x_{\I}) && \\
   & \ddots & \\
   & &  \hat{A}_{\mu}^{\I(T)}(x_{\I})
\end{array} \right),
\label{A_mu}
\end{eqnarray} 
where the diagonal block component, $\hat{A}_{\mu}^{\I(a)}$, 
is an $M^{(a)}\times M^{(a)}$ matrix and
all the off-diagonal block components vanish.
$\hat{A}_{\mu}^{\I(a)}$ are  
determined by
the flatness condition $F_{\mu\nu}^{\I}=0$, up to the gauge transformations
that are elements of $U(M^{(1)})\times \cdots \times U(M^{(T)})$ and
$y_{\I}$-independent.  The vacua on the total space
are parameterized by $M^{(a)}$, $\theta^{(a)}$ and 
$\hat{A}_{\mu}^{\I (a)}$ modulo 
the gauge transformations. 

Next, we examine the relation between the vacua on the total space and 
the base space. Each vacuum of (\ref{S_D}) is lifted up to a vacuum of 
(\ref{S_D+1}).
On the other hand, the configurations given by
(\ref{A_y}) and (\ref{A_mu}) are $y_{\I}$-independent,
so that they correspond to the vacua on the base space.
This implies that the map from the space of 
the vacua on the base space to those 
on the total space is surjective. However, it is not injective.
Suppose that $\hat{A}_{\mu}^{\I(a)}$ can be block-diagonalized as
\begin{eqnarray}
\hat{A}_{\mu}^{\I(a)}
=\left(\begin{array}{ccccc}
\ddots &&&& \\
& \hat{A}_{\mu}^{\I (a;s-1)} && \\
&& \hat{A}_{\mu}^{\I (a;s)} & \\
&&& \hat{A}_{\mu}^{\I (a;s+1)} & \\
&&&& \ddots  
\end{array} \right),
\label{hatA_mu}
\end{eqnarray} 
where $\hat{A}_{\mu}^{\I (a;s)}$ is an $N^{(a)}_s\times N^{(a)}_s$ matrix
and $M^{(a)}=\cdots+N^{(a)}_{s-1}+N^{(a)}_s+N^{(a)}_{s+1}+\cdots$.
Then, by applying the gauge transformation of the type $V_{\I}$, one can
shift $(\theta^{(a)},\cdots,\theta^{(a)})$ in $\hat{A}_y$ as 
\begin{align}
(\underbrace{\theta^{(a)},\cdots,\theta^{(a)}}_{M^{(a)}})
\rightarrow (\underbrace{\theta^{(a)},\cdots,\theta^{(a)}}_{M^{(a)}})
+(\cdots,
\underbrace{n^{(a)}_{s-1},\cdots,n^{(a)}_{s-1}}_{N^{(a)}_{s-1}},
\underbrace{n^{(a)}_{s},\cdots,n^{(a)}_{s}}_{N^{(a)}_{s}},
\underbrace{n^{(a)}_{s+1},\cdots,n^{(a)}_{s+1}}_{N^{(a)}_{s+1}},
\cdots),
\end{align}
where $n^{(a)}_s$ can be different.
We denote this shifted $\hat{A}_y$ by $\hat{A}_y'$.
The gauge-transformed configuration described by $\hat{A}_y'$ 
represents the same
vacuum on the total space as the original configuration described by
$\hat{A}_y$. As in the 
case of the trivial vacuum on the total space, due to 
the variety of the choices of $n^{(a)}_s$, one can consistently truncate the 
theory around the vacuum of (\ref{S_D+1}) described by (\ref{A_y}) and 
(\ref{A_mu}) in different ways to obtain different theories on the base space.
Those theories on the base space are the ones around the vacua of (\ref{S_D})
given by 
\begin{align}
&\hat{a}_{\alpha}^{\I(a;s)}
=-\frac{1}{R}(\theta^{(a)}+n^{(a)}_s)b_{\alpha}^{\I}\mathbf{1}_{N^{(a)}_s}
+e^{\I\mu}_{\alpha}\hat{A}_{\mu}^{\I(a;s)}, \nonumber\\
&\hat{\phi}=\hat{A}_y'.
\label{nontrivial vacua on base space}
\end{align}
Indeed, a general solution to the vacuum condition (\ref{vacuum condition})
takes this form in the gauge in which $\phi$ is diagonal.
It is seen from the first equation in (\ref{vacuum condition}) that
$e^{\I\mu}_{\alpha}\hat{A}_{\mu}^{\I(a;s)}$ gives a flat connection on
the base space.
It is easy to see that 
the T-duality also holds for the theories around the nontrivial vacua
on the total space.
In fact, by making $s$ in each block run from $-\infty$ to $\infty$,
taking $N^{(a)}_s=N^{(a)}$, $M^{(a)}=N^{(a)}\times\infty$ and $n^{(a)}_s=s$
and imposing the periodicity condition on the fluctuations around
(\ref{nontrivial vacua on base space}), one obtains
the theory around the vacuum of (\ref{S_D+1}) described by (\ref{A_y}) and 
(\ref{A_mu}) with $M^{(a)}$ replaced by $N^{(a)}$.

Finally, we comment on the quantization of the fluxes.
For the vacua (\ref{nontrivial vacua on base space}),
the first equation in (\ref{vacuum condition}) implies
\begin{align}
\hat{f}_{\alpha\beta}^{(a;s)}
=-\frac{1}{R}(\theta^{(a)}+n^{(a)}_s) b_{\alpha\beta}\mathbf{1}_{N^{(a)}_s}.
\label{fhat}
\end{align}
The 1st Chern class evaluated from both sides of (\ref{fhat}) 
is an element of the 2nd cohomology class with the integer coefficients of
the base manifold, $H^2(B,Z)$.
If the fiber bundle is nontrivial, 
the curvature of the principal $S^1$ bundle, $\frac{1}{R}b_{\alpha\beta}$, 
is a nontrivial element of 
$H^2(B,Z)$ so that we have a relation $p(\theta^{(a)}+n^{(a)}_s) \in Z$ with
a certain $p \in Z_+$.
This is the quantization of the fluxes. 
From this relation, we can deduce 
\begin{align}
\theta^{(a)}=\frac{l^{(a)}}{p},
\end{align}
where $l^{(a)}$ are integers satisfying $0 \leq l^{(a)} \leq p-1$.
If the fiber bundle is trivial, the curvature is a unit element of $H^2(B,Z)$
so that we have no relation for $\theta^{(a)}+n^{(a)}_s$.
There is no quantization of the fluxes in this case, and $\theta^{(a)}$
take continuous values.
For example, in $U(M)$ Yang Mills on $R^p\times S^1$, the vacua
are completely parameterized by $M^{(a)}$ and $\theta^{(a)}$
\cite{Hosotani:1983xw,Hosotani:1988bm}. Here
$\theta^{(a)}$ are continuous parameters and $\hat{A}_{\mu}=0$.
The value
of $p$ should also be determined
from the structure of the fundamental group on the 
total space,
because the vacua on the total space are given by the space of the flat
connections modulo the gauge transformations 
and the flat connections are classified by the holonomies
that are a representation of the fundamental group. 
One can see in the next section
that this is indeed the case in some examples.
Note that
in the $R^p\times S^1$ case, $\pi_1(R^p\times S^1)=\pi_1(S^1)=Z$ so that
there is no quantization of $\theta^{(a)}$.

\section{Examples}
\setcounter{equation}{0}
In this section, we present some examples of the T-duality for gauge theory
on curved space described as a
principal $S^1$ bundle. In sections 3.1, 3.2 and 3.3, we treat
$S^3$ and $S^3/Z_k$ as $S^1$ on $S^2$, $S^5$ and $S^5/Z_k$ as $S^1$ on $CP^2$ 
and the Heisenberg nilmanifold
as $S^1$ on $T^2$, respectively.

\subsection{$S^3$ and $S^3/Z_k$ as $S^1$ on $S^2$}
We consider $S^3$ with radius 2 and regard it as 
the $U(1)$ Hopf bundle on $S^2$ with radius 1. $S^3$ with radius 2 is 
defined by
\begin{eqnarray}
\{(w_1,w_2)\in C^2\:|\: |w_1|^2+|w_2|^2=4\}.
\label{definition of S^3}
\end{eqnarray}
The Hopf map $\pi:\:S^3 \rightarrow CP^1 \;(S^2)$ is defined by
\begin{eqnarray}
(w_1,w_2) \rightarrow [(w_1,w_2)]
\equiv \{\lambda(w_1,w_2)|\lambda \in C\backslash\{0\} \}.
\end{eqnarray}
Two patches are introduced on $CP^1$: the patch 1 $(w_1\neq 0)$ and the 
patch 2 $(w_2\neq 0)$. 
On the patch 1 the local trivialization is given by
\begin{eqnarray}
(w_1,w_2) \rightarrow \left(\frac{w_2}{w_1},\frac{w_1}{|w_1|}\right),
\label{local trivialization 1}
\end{eqnarray}
where $\frac{w_2}{w_1}$ is the local coordinate of $CP^1$, while
on the patch 2 the local trivialization is given by
\begin{eqnarray}
(w_1,w_2)\rightarrow \left(\frac{w_1}{w_2},\frac{w_2}{|w_2|}\right),
\label{local trivialization 2}
\end{eqnarray}
where $\frac{w_1}{w_2}$ is the local coordinate of $CP^1$.

The equation (\ref{definition of S^3}) is solved as 
\begin{eqnarray}
w_1=2 \cos\frac{\theta}{2}\: e^{i\sigma_1},\;\;\;
w_2=2 \sin\frac{\theta}{2}\: e^{i\sigma_2},
\end{eqnarray}
where $0 \leq \theta \leq \pi$ and $0 \leq \sigma_1,\:\sigma_2 < 2\pi$.
We put
\begin{eqnarray}
\varphi=\sigma_1-\sigma_2,\;\;\; \psi=\sigma_1+\sigma_2,
\end{eqnarray}
and can change the ranges of $\varphi$ and $\psi$ 
to $0\leq \varphi < 2\pi$ and $0 \leq \psi < 4\pi$.
The periodicity is expressed as 
\begin{eqnarray}
(\theta,\varphi,\psi)\sim (\theta,\varphi+2\pi,\psi+2\pi)
\sim (\theta,\varphi,\psi+4\pi).
\label{periodicity for angular variables}
\end{eqnarray}
$\mbox{}$From the local trivializations (\ref{local trivialization 1}) and 
(\ref{local trivialization 2}), one can see that 
$\theta$ and $\phi$ are regarded as the angular coordinates of the base space
$S^2$ through the stereographic projection. 
The patch 1 corresponds to $0\leq \theta <\pi$, while the patch 2 corresponds
to $0 < \theta \leq \pi$.
The metric of $S^3$ is given as follows:
\begin{align}
ds_{S^3}^2&=|dw_1|^2+|dw_2|^2 \nonumber\\
&=d\theta^2+\sin^2\theta d\varphi^2+(d\psi+\cos\theta d\varphi)^2.
\label{S3metric}
\end{align}

The correspondence to the notation of section 4 is as follows:
\begin{eqnarray}
&&z^M_{[1]}=(\theta,\varphi,\psi+\varphi), 
\;\;\;x^{\mu}_{[1]}=(\theta,\varphi),\;\;\;y_{[1]}=\psi+\varphi,
\nonumber\\
&&z^M_{[2]}=(\theta,\varphi,\psi-\varphi), 
\;\;\;x^{\mu}_{[2]}=(\theta,\varphi),\;\;\;y_{[2]}=\psi-\varphi,
\nonumber\\
&&b_{\theta}^{[1]}=0,\;\;\;b_{\varphi}^{[1]}=\cos\theta-1, \nonumber\\
&&b_{\theta}^{[2]}=0,\;\;\;b_{\varphi}^{[2]}=\cos\theta+1, \nonumber\\
&&R=2.
\end{eqnarray}
The metric and the zweibeins of the base space $S^2$ are given by
\begin{eqnarray}
&&ds_{S^2}^2=d\theta^2+\sin^2\theta d\varphi^2, \nonumber\\
&&e_{\theta}^1=1,\;\;\;e_{\varphi}^2=\sin\theta.
\label{S2metric}
\end{eqnarray}
(\ref{S_D}) takes the form
\begin{eqnarray}
S_{S^2}=\frac{1}{g_{S^2}^2}\int d\theta d\varphi \sin\theta
\mbox{Tr}
\left(\frac{1}{2}(f_{12}-\phi)^2+\frac{1}{2}(D_1\phi)^2+\frac{1}{2}(D_2\phi)^2 
\right).
\label{S_S^2}
\end{eqnarray}
The vacua of (\ref{S_S^2}) takes the form
\begin{eqnarray}
&&\hat{a}_{1}^{[1],[2]}=0, \nonumber\\
&&\hat{a}_{2}^{[1]}=\tan\frac{\theta}{2}\:\hat{\phi},\;\;\;
  \hat{a}_{2}^{[2]}=-\cot\frac{\theta}{2}\:\hat{\phi},\nonumber\\
&&\hat{\phi}=\frac{1}{2}\mbox{diag}(\cdots,n_{i-1},n_i,n_{i+1},\cdots).
\label{vacua of S_S^2}
\end{eqnarray}
The configuration of the $i$-th diagonal element of the gauge fields
are the Dirac monopole with the monopole charge $n_i/2$.
That $n_i$ are integers is consistent with Dirac's quantization condition.
The vacuum of Yang Mills on $S^3$ is unique due to $\pi_1(S^3)=0$.
There are no degrees of freedom corresponding to 
$\theta^{(a)}$ and $\hat{A}_{\mu}^{\I}$, and $p=1$.
The value of $p$ is also determined consistently by 
\begin{align}
W=P\exp\left(i\int_0^{4\pi} dy_{\I}\hat{A}_y\right)=1.
\end{align}
The theories around all the vacua (\ref{vacua of S_S^2}) originate from the
theory around the trivial vacuum on $S^3$.

As shown in the previous section in general, 
there holds the T-duality between  
the original $U(N)$  Yang Mills on $S^3$ and 
(\ref{S_S^2}) with the gauge group $U(N\times \infty)$ and
the periodicity condition imposed.
The relationship between the gauge fields on $S^3$ and the gauge fields
and the Higgs field on $S^2$ is given in (\ref{Fourier transf}) with
$\alpha=1,2$.
The gauge fields on $S^3$ are expanded in terms of the vector
spherical harmonics
on $S^3$, while the gauge fields and the Higgs field on $S^2$ in this case
are expanded together in terms of the vector monopole harmonics 
\cite{Wu:1976ge,Olsen:1990jm}.
$\mbox{}$From (\ref{Fourier transf}), one can read off
the relationship between the spherical harmonics on $S^3$ and the monopole
harmonics, which was found in \cite{ITT,ISTT} to show the T-duality between
$\mathcal{N}=4$ SYM on $R\times S^3(/Z_k)$ and 2+1 SYM on $R\times S^2$.

The lens space $S^3/Z_k$ $(k\in Z_+)$ 
is defined by introducing an identification,
$(w_1,w_2)\sim (w_1 \: e^{\frac{2\pi i}{k}}, w_2\: e^{\frac{2\pi i}{k}})$ 
into the definition of $S^3$ and can also be regarded as $S^1$ on $S^2$.
The local trivialization on the patch 1 is
\begin{eqnarray}
(w_1,w_2) 
\rightarrow \left(\frac{w_2}{w_1},\left(\frac{w_1}{|w_1|}\right)^k\right),
\end{eqnarray}
while the local trivialization on the patch 2 is 
\begin{eqnarray}
(w_1,w_2) 
\rightarrow \left(\frac{w_1}{w_2},\left(\frac{w_2}{|w_2|}\right)^k\right).
\end{eqnarray}
The radius of the fiber is replaced with $R=\frac{2}{k}$. The form of 
the action on the base space is the same as (\ref{S_S^2}). The counterpart
of (\ref{vacua of S_S^2}) is obtained by replacing $\hat{\phi}$
in (\ref{vacua of S_S^2}) with 
$\hat{\phi}=\frac{1}{2}\mbox{diag}
(\cdots,\beta_{i-1}+kn_{i-1},\beta_i+kn_i,\beta_{i+1}+kn_{i+1},\cdots)$,
where $\beta_i$ are integers and $0\leq \beta_i \leq k-1$.
The values and the multiplicities of $\beta_i$ label the vacua on $S^3/Z_k$
and correspond to $\theta^{(a)}$ and $M^{(a)}$, respectively.
The vacua on $S^3/Z_k$ are
classified by the holonomy along the generator of $\pi_1(S^3/Z_k)=Z_k$,
which can be evaluated from
\begin{align}
W=P\exp \left(i\int_0^{4\pi/k}dy_{\I}\hat{A}_y\right).
\end{align}
The value of $p$ 
is determined by $W^k=1$ as $p=k$.

\subsection{$S^5$ and $S^5/Z_k$ as $S^1$ on $CP^2$}
We regard $S^5$ as a $U(1)$ bundle on $CP^2$ as follows.
$S^5$ with the radius 1 is defined by
\begin{eqnarray}
\{(w_1,w_2,w_3)\in C^3\:|\:|w_1|^2+|w_2|^2+|w_3|^2=1\}.
\label{definition of S^5}
\end{eqnarray}
The Hopf map $\pi:\:S^5\rightarrow CP^2$ is given by
\begin{eqnarray}
(w_1,w_2,w_3) \rightarrow [(w_1,w_2,w_3)]
\equiv \{\lambda(w_1,w_2,w_3)|\lambda \in C\backslash\{0\} \}.
\end{eqnarray}
Three patches are introduced on $CP^2$: the patch 1 $(w_1\neq 0)$, the 
patch 2 $(w_2\neq 0)$ and the patch 3 $(w_3\neq 0)$.
The fiber on the patch $I$ given by the local trivialization
is parameterized by $\frac{w_{I}}{|w_{I}|}$.
(\ref{definition of S^5}) is solved as
\begin{eqnarray}
&&w_1=\cos\chi \: e^{i\tau}, \nonumber\\
&&w_2=\sin\chi\cos\frac{\theta}{2} \: e^{i(\tau+\frac{\psi+\varphi}{2})}, 
\nonumber\\
&&w_3=\sin\chi\sin\frac{\theta}{2} \: e^{i(\tau+\frac{\psi-\varphi}{2})}.
\end{eqnarray}
where $0\leq\tau < 2\pi$, $0\leq \chi \leq \frac{\pi}{2}$, 
$0\leq\theta\leq\pi$, $0\leq\varphi < 2\pi$ and $0\leq\psi < 4\pi$.
The periodicity in this case is given by
\begin{eqnarray}
(\chi,\theta,\varphi,\psi,\tau)\sim (\chi,\theta,\varphi,\psi,\tau+2\pi)
\sim (\chi,\theta,\varphi,\psi+4\pi,\tau)
\sim (\chi,\theta,\varphi+2\pi,\psi+2\pi,\tau).
\label{periodicity for S^5}
\end{eqnarray}
The metric of $S^5$ is given by
\begin{align}
ds_{S^5}^2 &= |dw_1|^2+|dw_2|^2+|dw_3|^2 \nonumber\\
&= ds_{CP^2}^2 + \omega^2 
\label{metric of S^5}.
\end{align}
with 
\begin{align}
&ds_{CP^2}^2=d\chi^2+\frac{1}{4}\sin^2\chi(d\theta^2+\sin^2\theta d\varphi^2
+\cos^2\chi(d\psi+\cos\theta d\varphi)^2), \nonumber\\
&\omega=d\tau+\frac{1}{2}\sin^2\chi(d\psi+\cos\theta d\varphi),
\end{align}
where $ds_{CP^2}^2$ is the metric of $CP^2$, which is called the
Fubini-Study metric
while $\omega$ is the connection 1-form.
The correspondence to the notation of section 4 is
\begin{eqnarray}
&&z^M_{[1]}=(\chi,\theta,\varphi,\psi,\tau), \;\;\;
x^{\mu}_{[1]}=(\chi,\theta,\varphi,\psi),\;\;\;y_{[1]}=\tau,
\nonumber\\
&&z^M_{[2]}=(\chi,\theta,\varphi,\psi,\tau+\frac{1}{2}(\psi+\varphi)), \;\;\;
x^{\mu}_{[2]}=(\chi,\theta,\varphi,\psi),\;\;\;
y_{[2]}=\tau+\frac{1}{2}(\psi+\varphi),
\nonumber\\
&&z^M_{[3]}=(\chi,\theta,\varphi,\psi,\tau+\frac{1}{2}(\psi-\varphi)), \;\;\;
x^{\mu}_{[3]}=(\chi,\theta,\varphi,\psi),\;\;\;
y_{[3]}=\tau+\frac{1}{2}(\psi-\varphi),
\nonumber\\
&&b_{\chi}^{[1],[2],[3]}=0,\;\;\;b_{\theta}^{[1],[2],[3]}=0, \nonumber\\
&&b_{\varphi}^{[1]}=\frac{1}{2}\sin^2\chi\cos\theta, \;\;\;
b_{\psi}^{[1]}=\frac{1}{2}\sin^2\chi, \nonumber\\
&&b_{\varphi}^{[2]}=\frac{1}{2}(\sin^2\chi\cos\theta-1), \;\;\;
b_{\psi}^{[2]}=\frac{1}{2}(\sin^2\chi-1), \nonumber\\
&&b_{\varphi}^{[3]}=\frac{1}{2}(\sin^2\chi\cos\theta+1), \;\;\;
b_{\psi}^{[3]}=\frac{1}{2}(\sin^2\chi-1), \nonumber\\
&&R=1.
\label{S^5 correspondence}
\end{eqnarray}
$b_{\mu}^{\I}$ in (\ref{S^5 correspondence}) is 
called the gravitational and electromagnetic instanton in 
\cite{Trautman:1977im}.
The vacuum on $S^5$ is unique. The vacua on $CP^2$ are given by
(\ref{nontrivial vacuum}). 
The value of $p$ is determined by 
$W=P\exp \left(i\int_0^{2\pi}dy_{\I}A_y\right)=1$ as $p=1$.

The lens space $S^5/Z_k$ is treated in the same way as $S^3/Z_k$.

\subsection{Heisenberg nilmanifold as $S^1$ on $T^2$}
The Heisenberg nilmanifold \cite{Kachru:2002sk,Bouwknegt:2003vb} 
is a twisted 3-torus 
that has the following 
periodicity condition.
\begin{eqnarray}
(x^1,x^2,y) 
\sim (x^1,x^2+L_2,y) 
\sim (x^1,x^2,y+L_y) 
\sim (x^1+L_1,x^2,y-\kappa L_1x^2),
\label{periodicity of nilmanifold}
\end{eqnarray}
where $x^1$, $x^2$ and $y$ are the coordinates of the nilmanifold, and 
$\kappa$ is determined from consistency of 
(\ref{periodicity of nilmanifold}) as 
$\kappa=l\frac{L_y}{L_1L_2},(l \in Z)$. 
The metric of the nilmanifold is given by
\begin{eqnarray}
ds^2=(dx^1)^2+(dx^2)^2+(dy+\kappa x^1dx^2)^2.
\end{eqnarray}

We regard the nilmanifold as a $U(1)$ bundle on $T^2$ parameterized 
by $x^1$ and $x^2$. 
We need two patches on $T^2$ to describe the $U(1)$ bundle.
We define the patch 1 as the region 
$0 < x^1_{[1]} < L_1$, and the patch 2 as the region
$-\frac{L_1}{2} < x^1_{[2]} < \frac{L_1}{2}$.
On each patch, the nilmanifold is locally trivialized  
such that it is parameterized by $(x^1_{[1]},x^2_{[1]},y_{[1]})$ 
on the patch 1 and $(x^1_{[2]},x^2_{[2]},y_{[2]})$ on the patch 2, where 
$(x^1_{\I},x^2_{\I})$ are the local coordinate of the base manifold
$T^2$ and $y_{\I}$ parameterizes the $S^1$ fiber direction. 
On the overlap between the two patches, the transition functions are given as 
follows:
\begin{eqnarray}
x^1_{[2]}=x^1_{[1]}\;\;,x^2_{[2]}=x^2_{[1]},\;\;y_{[2]}=y_{[1]} ,
\hspace{3mm}
\end{eqnarray}
in the region 
$0<x^1_{[1]}<\frac{L_1}{2},\;0< x^1_{[2]}<\frac{L_1}{2}$, and  
\begin{eqnarray}
x^1_{[2]}=x^1_{[1]}-L_1,\;\;x^2_{[2]}=x^2_{[1]},\;\;
y_{[2]}=y_{[1]}-\kappa L_1x^2_{[1]},
\end{eqnarray}
in the region  
$\frac{L_1}{2}<x^1_{[1]}<L_1,\;-\frac{L_1}{2}<x^1_{[2]}<0$.
Note that this representation of the nilmanifold in terms of the patches
is equivalent to the definition in terms of the periodicity 
condition (\ref{periodicity of nilmanifold}).

The correspondence to the notation of section 4 is as follows:
\begin{eqnarray}
&&b_1^{\I}=0,\;\; b_2^{\I}=\kappa x^1_{\I}, \;\; I=1,2, \nonumber\\
&&b_{12}=\kappa, \nonumber\\
&&R=\frac{L_y}{2\pi}.
\end{eqnarray}
$b_{\alpha}^{\I}$ represents the constant magnetic flux with
the strength $l$ on $T^2$. This implies $p=l$.
The value of $p$ is also determined from 
the structure of the fundamental group of the Heisenberg nilmanifold.
As discussed in \cite{Kachru:2002sk}, $W^l$ equals the Wilson line
along an element of the 
commutator subgroup of the fundamental group.
For the $U(1)$ part of the $a$-th block of
the gauge fields, therefore, we have
\begin{align}
\left(P\exp\left(i\int_0^{L_y}dy_{\I}\hat{A}_y^{(a)}\right)\right)^l=1,
\end{align}
from which $p=l$ follows.
In this case, $e^{\I\mu}_{\alpha}\hat{A}_{\mu}^{\I(a;s)}$ in 
(\ref{nontrivial vacua on base space}) can give nontrivial Wilson lines
along the generators of the fundamental group of $T^2$
and contribute to the classification of the vacua.
(\ref{S_D}) takes the form
\begin{eqnarray}
S=\frac{1}{g_{T^2}}\int dx_1 dx_2 {\rm{Tr}}\left(
\frac{1}{2}(f_{12}+\kappa\phi)^2+\frac{1}{2}(D_1\phi)^2
+\frac{1}{2}(D_2\phi)^2\right).
\label{action on T^2}
\end{eqnarray}

\section{Conclusion and discussion}
\setcounter{equation}{0}
In this paper, we first discussed the variety of the consistent truncations 
from Yang Mills on the total space to Yang Mills with the Higgs field 
on the base space in the trivial and nontrivial principal $S^1$ bundles.
Different consistent truncations of the theory around a vacuum 
of Yang Mills on the total space
yield the theories around different 
nontrivial vacua of Yang Mills with the Higgs
field on the base space. In the case of the nontrivial $S^1$ bundles, 
the nontrivial vacua on the base space have monopole-like gauge configurations.
By using this viewpoint, we showed the T-duality between 
the theories on the total space and the base space in the nontrivial bundle
case as well as the trivial bundle case.
The difference between these two cases is that
in the nontrivial bundle case, the vacuum configurations of the gauge fields
are the monopole-like ones and 
the Fourier transformation must be made locally on each patch. 
It is remarkable that
the monopole-like charges are identified with the momenta on the total space.
We also classified the vacua on the total space and their relation to the 
vacua on the base space.
The quantization of the monopole-like charges on the base space is 
understood from the structure of the fundamental group on the total space.

It is easy to add adjoint matters to Yang Mills on the total space and to
introduce supersymmetry. In \cite{ISTT}, we showed the T-duality between
the theory around the trivial vacuum of $\mathcal{N}=4$ SYM on 
$R\times S^3/Z_k$ and the theory around a vacuum of 2+1 SYM on $R\times S^2$.
Our results in this paper show that the T-duality also holds for the
nontrivial vacua of $\mathcal{N}=4$ SYM on 
$R\times S^3/Z_k$. 

In this paper, we restricted ourselves to principal $U(1)$ bundles. Nonabelian
generalization is important. Typical examples are $S^7$ as $S^3\;(SU(2))$ on
$S^4$, $SU(3)$ as $U(2)$ on $CP^2$ and so on.
In \cite{ISTT}, we showed that the theory around each vacuum of Yang Mills with
the Higgs on $S^2$ is equivalent to the theory around a vacuum described by 
fuzzy spheres of a matrix model. This means that
Yang Mills on $S^3$ and $S^3/Z_k$ is realized
in the matrix model. It is interesting to examine what condition is needed
in order for a gauge theory on a fiber
bundle to be realized in a matrix model.
Finally, we expect to apply our findings to (flux) compactification in 
string theory.

\section*{Acknowledgements}
We would like to thank 
H. Kawai for stimulating discussions.
The work of A.T. is supported in part by Grant-in-Aid for Scientific
Research (No.16740144) from the Ministry of Education, Culture, Sports,
Science and Technology.




\end{document}